\documentstyle[11pt,moriond,epsfig]{article}

\def\PRL{\em Phys. Rev. Lett.}
\def\PRD{{\em Phys. Rev.} D}

\def\AA{\em Astron. Astrophys.}

\def\be{\begin{equation}}
\def\ee{\end{equation}}
\def\bea{\begin{eqnarray}}
\def\eea{\end{eqnarray}}

\begin{document}
\vspace*{4cm}
\title{TRIGGERS FOR THE DETECTION OF GRAVITATIONAL WAVE BURSTS}

\author{N. ARNAUD, F. CAVALIER, M. DAVIER, \underline{ P. HELLO}
 AND T. PRADIER}

\address{Laboratoire de l'Acc\'el\'erateur Lin\'eaire\\
B.P. 34\\
B\^atiment 208, Campus d'Orsay\\
91898 Orsay Cedex, France}

\maketitle\abstracts{We present several filtering methods which can be used as triggers for the detection of 
gravitational wave bursts in interferometric detectors. All the methods are compared to matched filtering with
the help of a {\it figure of merit} based on the detection of  supernovae signals simulated by 
Zwerger and M\"uller.}

\section{Introduction}

Supernovae have been historically the first envisaged source of gravitational waves (GW). 
Although binary inspirals or even periodic
GW emitters like pulsars seem to be nowadays more promising sources, impulsive sources 
of GW such as supernovae should also be considered
in the data analysis design of interferometric detectors currently under construction.

Impulsive GW sources are typically collapses of massive stars, leading to the 
birth of a neutron star 
(type II supernova) \cite{bona,zwerger,rampp} or of a black hole \cite{stark}; 
mergers of compact binaries can also be considered as impulsive sources \cite{ruffert}. 

The problem with these sources is that the emitted waveforms are very poorly predicted, 
unlike the binary inspirals.
As a consequence, this forbids the use of matched filtering for the detection of GW 
bursts in the data of one interferometric detector.
The filtering of such bursts should therefore be as general and robust as possible and with 
minimal {\it a priori} assumptions on the waveforms to be detected. A drawback
is of course that such filters will be sensitive to non-stationary noises as well as to GW bursts;
spurious events, e.g. generated by these transient noises, should be eliminated afterwards when
working in coincidence with other detectors.
But, on the other hand, burst filters could help to identify and understand these noises, which would
be useful especially during the debugging  phase of the detector.

We present in the following some filtering methods dedicated to the detection of GW bursts : methods based
on the autocorrelation, slope detector, correlator ... All the filters are compared by studying their performance
to detect a reference sample of GW burst signals; for this purpose, just as in \cite{nous} (and in order to use somewhat physically sound signals),
we use the catalogue of signals emitted by type II supernovae, numerically computed by Zwerger and M\"uller 
(ZM)\cite{zwerger} and available on the web \cite{webSN}.

Throughout the following, we assume that the detector noise 
is white, stationary and Gaussian with zero mean. For the numerical 
estimates, we chose the flat (amplitude) spectral density to 
be $h_n \simeq 4\times10^{-23} /\sqrt{\mathrm Hz}$ and the sampling 
frequency $f_s \simeq 20$ kHz, so the standard deviation of the noise is
$\sigma_n = h_n \sqrt{f_s/2} \sim 3\times10^{-21}$; we will note the 
sampling time $t_s=1/f_s$. The value chosen for $h_n$ corresponds 
approximately to the minimum of the sensitivity curve of the 
VIRGO detector \cite{virgosens};
around this minimum, the sensitivity is rather flat, in the range $\sim$ 
[200 Hz,1kHz], which is precisely the range 
of interest for the gravitational wave bursts
we are interested in. This validates then our assumption of a white noise; 
otherwise, we can always assume that the detector output
 has been first whitened by a suitable filter \cite{cuo}.

\section{General filters}

\subsection{Filters based on the autocorrelation}

The noise being whitened, its autocorrelation is ideally a Dirac function, and, in
practice vanishes outside of 0. The autocorrelation of the data $x(t)$
\begin{equation}
A_x(\tau) = \int x(t) x(t+\tau) dt
\end{equation}
should then reveal
the presence of some signal (surely correlated). The information contained in the autocorrelation
function can be extracted in different ways. We have studied two of them and built so two non-linear 
filters. The first one computes the maximum of the autocorrelation $A_x(\tau)$; this occurs always at $\tau=0$, and then
this maximum is nothing but the norm of $x(t)$. For sampled data $x_i$ in a window of size $N$, the output of this
filter is simply
\begin{equation}
A(0) = \sum_{i=1}^N x_i^2.
\end{equation}
In the following, we will refer to this filter as the Norm Filter (NF). A similar approach has been developed independently
by Flanagan and Hughes in the context of the detection of binary black hole mergers \cite{flana}.

Another simple possibility is to look at the norm of the autocorrelation function :
\begin{equation}
||A|| = \sqrt{ {1 \over N} \sum_{k=2}^N A(k)^2},
\end{equation}
where $A(k)$ denotes the discrete autocorrelation of $N$ data $x_i$. The sum is here initiated at the second bin according to the fact
that the noise (uncorrelated) contributes essentially to the first bin. In the following we will call this filter Norm of 
Autocorrelation (NA). In practice, the $A(k)$ are computed in the Fourier space, according to the Wiener-Khintchine theorem,  allowing the use
of FFT's.
Note that the only parameter for these two filters is the window size $N$.

\subsection{The Bin Counting filter}

This filter (BC) computes the number of bins in a window of size N whose value exceeds some threshold $s \times \sigma_n$. For example,
if we take $s=2$ and pure Gaussian noise, as $P(|x_i| \geq 2\sigma_n) \simeq 4.6\%$, the output of the BC filter is on average about
46 'counts' for a window size $N=1000$. This filter is also non-linear, but it involves two parameters : the window size $N$
and the threshold $s$. The threshold $s$ is chosen by maximizing the signal to noise ratio (SNR) when detecting the signals of the ZM catalogue.
The optimum is for $s \simeq 1.7$ but it is not critical; indeed any value of $s$ in the range [1.4,2.0] would be also convenient (with a low
loss in SNR).

\subsection{The Slope Detector}

This filter (SD) fits the data in a window of size N to a straight line. If the data are pure white noise with zero mean, then the slope
of the fitted line is zero on average, so this slope detector can well discriminate between the two cases : only noise or noise+signal.
The output of the SD is simply
\begin{equation}
a = { <tx> - <t><x> \over <t^2> - <t>^2} = {1 \over N }\sum_{i=1}^N { t_i-<t> \over <t^2> - <t>^2} x_i,
\label{slope}
\end{equation}
where $<y>=\sum_{i=1}^N y_i/N$ denotes the mean value of the $y_i$ and $t_i= i\times t_s$. Note that this filter is linear,
as opposed to the first three considered. Again the only parameter is the window size $N$.

\subsection{The Peak Correlator}

Filtering by correlating the data with peak (or pulse) templates is justified by the fact that simulated supernova GW signals exhibit one
(or more) peaks. The pulse templates have been built from truncated Gaussian functions 
\begin{equation}
F_\tau(t) = \exp\left( - {t^2 \over 2 \tau}\right),
\end{equation}
with $-3\tau \leq t \leq 3\tau$. The only parameter of the peak correlator (PC) is the width of the Gaussian pulse filter $\tau$ (the window size
is automatically set to be a power of 2, due to use of FFT's). The lattice of filters is then built as usual (see \cite{sathya} for example) :
the distance $\Delta\tau$ between two successive filters of the lattice $F_\tau$ and $F_{\tau+\Delta\tau}$ is computed by the condition
\begin{equation}
{<F_\tau,F_\tau> - <F_{\tau+\Delta\tau},F_\tau> \over 
<F_\tau,F_\tau>} \leq \epsilon,
\end{equation}
where we define a scalar product as $<f,g> = {\mathrm Max }_{t'}\,\,{\int f(t+t') g(t) dt 
/  \sqrt{\int f^2(t)dt}}$ and $\epsilon$ is the allowed loss in the SNR. A simple calculation leads
to $\Delta\tau=2\tau\sqrt{\epsilon}$. With $\epsilon=10^{-2}$, we finally have 26 templates in the interval
[0.1 ms, 10 ms] (which are used in the following).

\subsection{Statistics}

The SD and PC filters being linear, they transform an input normal Gaussian noise with zero mean into a 
Gaussian noise with zero mean
but with a modified standard deviation. For the SD filter, with the help of Eq.\ref{slope}, we find a standard deviation
\begin{equation}
\sigma_{\mathrm SD}^2 = \sum_{i=1}^N \left( {t_i - <t> \over N(<t^2> - <t>^2)} \right)^2 = {12 f_s^2 \over N (N^2-1)}.
\end{equation}
Similarly, when correlating pure noise data with the pulse filter $F_\tau$, we obtain a Gaussian noise of standard deviation
\begin{equation}
\sigma_{\mathrm PC}^2 = \sqrt{\pi} \, { \tau \over t_s}.
\end{equation}

The output of the BC filter is a binomial random variable, considering the data are pure Gaussian random variables;
it  is well approximated by Gaussian statistics for long enough windows (typically $N\geq 50$) \cite{stat} and 
the standard deviation for the noise at the BC output is simply 
\begin{equation}
\sigma_{\mathrm BC} = \sqrt{Np(1-p)}
\end{equation}
with $p={\mathrm erfc}(s/\sqrt{2})$.

Considering the NF filter, if we call $A$ its output, it is easy to see that $A$ follows a chi-square distribution with $N$ degrees of freedom,
and then $A^* = \sqrt{2A}-\sqrt{2N-1}$ is also well approximated by a normal
 ($\sigma_{\mathrm NA}=1$) Gaussian
random variable, provided $N\geq 30$ \cite{stat}, and the input noise is itself a normal Gaussian
random variable.

Finally, the noise at the output of the NA filter is not known analytically and its characteristics have to be found numerically (adding some complexity
to this filtering method).

\section{Performance of the filters}

\subsection{Definition of a threshold for detection}

We set the  false alarm rate for each of the filters to be $10^{-6}$ (72 false alarms / hour for a sampling
rate $f_s=20$ kHz). This corresponds to a detection threshold (normalised SNR) of  $\eta \simeq 4.89$ for a single
'Gaussian filter'. For a trigger that incorporate in fact several filters, for example the 26 templates of the PC, the threshold
has to be raised accordingly, in order to keep a global false alarm rate of $10^{-6}$ (e.g. $\eta \simeq 5.50$ in this case).

\subsection{The Zwerger and M\"uller Catalogue}

The catalogue of Zwerger and M\"uller \cite{webSN} (ZMC) contains 78 
gravitational-wave signals. Each of them corresponds to a particular 
set of parameters, essentially the initial distribution of angular momentum 
and the rotational energy of the star core,
in the axisymmetric collapse models of ZM. The  signal total durations 
range from about 40 ms to a little more than 200 ms. 
The gravitational wave amplitudes of the stronger signals are of the order 
$h \sim$ a few $10^{-23}$ for a source located at 10\,Mpc. All the signals are computed for a source located at 10\,Mpc.
We can then re-scale the waveforms in order to locate the source at any distance $d$, according to
\begin{equation}
h(d,t) = h_0(t) \,\,{10 {\mathrm Mpc} \over d}
\end{equation}
where $h_0$ is the signal at 10\,Mpc and $h(d,t)$ is the same signal but at a distance $d$.
Concerning the shape of the waveforms , 
Zwerger and M\"uller distinguish three different types of signals\cite{zwerger}. 
Type I signals typically present 
a first peak (associated to the bounce)followed
by a ringdown. Type II signals show a few (2-3) decreasing peaks, 
with a time lag between the first two  of at least 10\,ms. Type III signals exhibit
no strong peak but fast ($\sim$ 1 kHz) oscillations after the bounce. 

Since the 78 signal waveforms are known, we can explicitly derive the 
optimal SNR provided by the Wiener filter matched to each of them, and 
then compute the
maximal distance of detection. We will then be able to build a benchmark 
for the different filters by comparing their results (detection distances) to the results
of the Wiener filter. Note that, in what follows, we consider optimally polarized GW's, along
the interferometer arms.

Let's call $h(t)$ one of the 78 signals (at some distance $d$) and $\tilde{h}(f)$ its Fourier transform. 
The optimal signal to noise ratio $\rho_0$ is given by 
\begin{equation}
\rho_0^2 = 2 \int {|\tilde{h}(f)|^2 \over S_h(f)} df= { f_s \over \sigma_n^2} \int |\tilde{h}(f)|^2  df 
= { f_s \over \sigma_n^2} \int |{h}(t)|^2  dt,
\end{equation}
where $S_h=h_n^2$ is the one-sided noise power spectral density (hence the factor of 2). 

\begin{figure}[ht]
\begin{center}
\epsfig{file=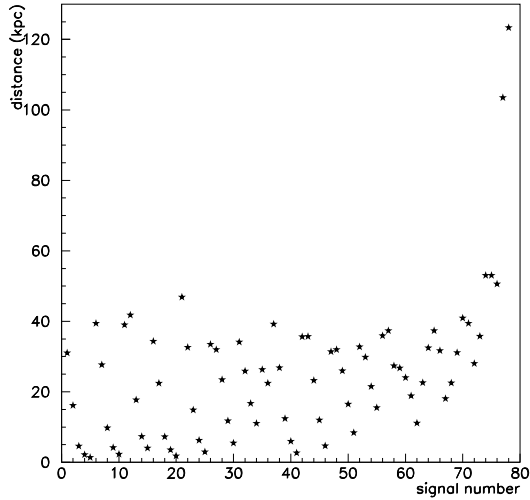, height=3.0in}
\caption{Detection distances calculated with the optimal filter for the 78
signals in the ZMC}
\label{optdist}
\end{center}
\end{figure}

As previously, a supernova signal is detected by the Wiener filter 
if $\rho_0 \geq \eta$, where $\eta$ is the same detection threshold as defined above.
Fig.\ref{optdist} shows the maximal distance of detection for each of the 78 
signals. The mean distance, averaged over all the signals, is about $\bar{d}_{\mathrm opt} \simeq 25.4$\,kpc, 
which is of the order of the diameter of the Milky Way. 
A  few signals can be detected at distances beyond 50\,kpc, 
the distance of the Large Magellanic Cloud (LMC). 
It is clear that this class of signals will be detected by the first generation interferometric 
detectors only if the supernovae occur inside our Galaxy or in the very
close neighbourhood.

\subsection{Estimating  a filter performance}

Let's consider one signal, say the $i^{\mathrm th}$ in the ZMC. The optimal filtering
allows to detect such a signal for a source located at the distance $d_i^{\,(0)}$. Similarly,
a filter F is able to detect the same signal up to a distance $d_i$; of course $d_i$ is averaged
over many noise realizations in a Monte Carlo simulation. The detection efficiency of the filter
F for this signal $i$ is simply defined as the distance of detection relative to
the optimal distance of detection : $d_i/d_i^{\,(0)}$. The global performance of the filter
F is then estimated as the detection efficiency averaged over all the signals of the ZMC :
\begin{equation}
\rho= {1 \over 78} \sum_{i=1}^{78} {d_i \over d_i^{\,(0)}}.
\end{equation}

\subsection{Comparison of the filtering methods}

The results for the different filters are reported in the Table \ref{tab} below.
We also give the average distance of 
detection $\bar{d}= {1 \over 78} \sum_{i=1}^{78} d_F^{(i)}$ for all the filters, together with 
the ratio $\bar{d}/\bar{d}_{\mathrm opt}$.

\begin{table}[ht]
\caption{Efficiency of the different filters. L means linear filter and NL means
non-linear filter.\label{tab}}
\vspace{0.4cm}
\begin{center}
\begin{tabular}{|c|c|c|c|c|c|c|}
\hline
Filter& Optimal & NF& NA & BC & SF &PC \\ 
\hline
$\bar{d}$ (kpc)& 25.4 & 11.5 & 11.4 & 10.9 & 20.7 & 18.5\\
$\bar{d}/\bar{d}_{\mathrm opt}$ & 1 & 0.45 & 0.45 & 0.43 &0.81& 0.73\\
$\rho$                 & 1      &0.46&0.47&0.43& 0.79 &  0.73\\
Linearity &L&NL&NL&NL&L&L \\ 
\hline
\end{tabular}
\end{center}
\end{table}

The three first filters NF, NA and BC (all non-linear) have an efficiency a little less than one half, while the SD and the PC
have an efficiency a little above 0.7. note that the SF has been in fact implemented with a sampling of 6 different window sizes,
sufficient to cover the variety of signals. If implemented with a single window size, as the other filters NF, NA and BC,
its performance decreases down to about 0.6.

\section{Conclusion}

We have discussed several filters to be used as triggers for detecting GW bursts in interferometric detectors.
They are all sub-optimal but their efficiency is not far below that of optimal filter.

Concerning the detection of ZMC like signals, we note that
none of the BC, NF and NA filters is efficient enough to cover  the whole Galaxy in average, at the contrary of the SD and PC 
(and optimal) filters.
Several signals can be 'seen' in fact anywhere from the Galaxy and even beyond; in particular the signals 77 and 78 can be 
detected up to the LMC by any of the filters. 

Finally, all the filters studied here can be implemented on line without problem, due to use of FFT's (for the NA and the PC)
or to simple recurrence relations between filter outputs in successive windows (NF,BC or SD).

More information (preprints, Virgo reports ...) can be found at \cite{monweb}

\end{document}